\documentclass[12pt]{article}
\usepackage{latexsym}
\usepackage{amssymb}
\usepackage[dvips]{graphicx}
\usepackage{epsfig}
\usepackage{rotating}

\usepackage{cancel}

\usepackage{amsfonts}
\usepackage{amsmath}

\newcommand{\bea}{\begin{eqnarray}}
\newcommand{\eea}{\end{eqnarray}}

\makeatletter

\begin{document}

\begin{titlepage}
\begin{flushright}
pi-other-246\\
ICMPA-MPA/2011/009\\
\end{flushright}

\begin{center}

{\Large\bf
Coherent states  for continuous spectrum operators \\
\medskip
with non-normalizable fiducial states
}

\bigskip

Joseph Ben Geloun$^{a,c,*}$,
Jeff Hnybida$^{a,\dag}$ and John R. Klauder$^{b,\ddag}$

\bigskip

$^a${\em Perimeter Institute for Theoretical Physics, 31 Caroline
St N} \\
{\em ON N2L 2Y5, Waterloo, ON, Canada} \\
\medskip
$^{b}${\em Department of Physics and
Department of Mathematics}\\
{\em University of Florida, Gainesville, FL 32611-8440}\\
\medskip
$^{c}${\em International Chair in Mathematical Physics
and Applications}\\
{\em ICMPA--UNESCO Chair, 072 B.P. 50  Cotonou, Republic of Benin} \\
\medskip
E-mails:  $^{*}$jbengeloun@perimeterinstitute.ca,
$^{\dag}$jhnybida@perimeterinstitute.ca,
\quad $^{\ddag}$klauder@phys.ufl.edu

\begin{abstract}
The problem of building coherent states from non-normalizable
fiducial states is considered. We propose a way of constructing
such coherent states by regularizing the divergence
of the fiducial state norm. Then, we successfully apply the
formalism to particular cases involving systems with
a continuous spectrum: coherent states
for the free particle and for the inverted oscillator $(p^2 - x^2)$
are explicitly provided. Similar ideas can be used
for other systems having non-normalizable fiducial states.
\end{abstract}

\end{center}

\noindent Pacs numbers: 03.65.-w

\end{titlepage}

 %\maketitle

%\setcounter{footnote}{0}

\section{Introduction}
\label{sect:intro}

Coherent states are acknowledged to be essential objects in understanding the
connection between the quantum and the classical counterparts of a system \cite{kl}-\cite{klish}.
Furthermore, they have become of significance in more
formal and interesting mathematical developments \cite{kbs}\cite{perel}\cite{od}\cite{al}.
In fact, coherent states can be constructed for any formal Hilbert space, having either a discrete
or continuous basis, including those related to, for instance, a quantum
mechanical system \cite{gk}.
Note that the problem of continuous spectrum dynamical operators
and the construction of their associated coherent states
have been addressed in different fashions but, generally
most can be recast in
the formulation of Gazeau-Klauder (GK) \cite{gk}.
Besides, in the latter work, the authors have provided
an axiomatic approach for families of states in order
to be called {\it physical} coherent states.
The GK axioms are simple to describe: continuity in the labels, a resolution of the
identity of the Hilbert space, temporal stability
of states and the action identity.
This approach has led to other developments on continuous spectrum
operators. For instance, it turns out that  ladder operators
for continuous spectrum operators can be defined
and corresponding eigenvalue problems for defining coherent states
can be explicitly solved \cite{kbg}. Statistical properties
and other nonclassical properties of such states have been recently
addressed \cite{new}.

Coherent states may be constructed in three different
ways: they can be solutions of an eigenvalue problem of an annihilation operator; they can be constrained to satisfy some uncertainty
principle or, finally, they can be defined as orbits of a unitary
operator acting on a preferred or {\it fiducial} state.
In the continuous spectrum case, it might happen that the eigenstates of the Hamiltonian are all non-normalizable. Focusing on the last proposal,
the family of states generated by a unitary transformation of any
non-normalizable fiducial state will be obviously non-normalizable
and so  would not satisfy elementary properties
of coherent states. Clearly, the problem of non-normalizable
states is recurrent when one is dealing with eigenstates of continuous spectrum
operators. Building coherent states for non-normalizable fiducial states has
 already been considered in the literature
and solved with different techniques according to
the particular model under consideration \cite{bart,jannus,flo,klish},
and it still attracts attention \cite{jose}.
We provide, in the present work, a specific formulation to address this issue of non-normalizable fiducial states and how to
define coherent states based on them.

In this paper, we consider the construction of coherent states
for
dynamical systems with continuous spectrum having non-normalizable states.
Two ways of regularizing the fiducial state are introduced
and, for these choices, two families of coherent states can be
built in the canonical quantization framework. We also investigate
the main properties of the coherent states with respect to the
GK axioms. We find that the two coherent state families are indeed
continuous in labels, normalized and satisfy a resolution
of the identity on the Hilbert space.
The remaining GK axioms has to be checked case
by case. Explicit examples are treated for the
free particle coherent states and the so-called inverted
harmonic oscillator \cite{bart}.
Interestingly, we show that in a leading approximation
in the regularization parameter, coherent states of these two
examples are stable under time evolution.

The paper is organized as follows: the next section
is devoted to the general formulation of canonical
coherent states with the regularized fiducial states.
Section \ref{sect:apps} consists in the applications
of our formalism: first the free particle
and, then, the inverted harmonic oscillator are investigated.
Section \ref{concl} gives a summary of our results
and an appendix provides useful identities and other
details for the text.

\section{Coherent states and non-normalizable fiducial
states}
\label{sect:cs}

In the present section, we formally discuss
a way of constructing coherent states based on a non-normalizable fiducial
vector. This vector is  selected among the eigenbasis spanning the Hilbert space related with the spectral decomposition of a given Hamiltonian operator.
Although our study is tailored to operators of the continuous spectrum type,
this does not preclude a natural extension of the formulation to the continuous and discrete spectrum situations.

Let us consider a Hamiltonian operator $H$ with a  continuous energy spectrum and let $\{|\Psi_E\rangle\}$ be the family
of eigenvectors  of $H$, i.e.
\bea
H |\Psi_E\rangle =E |\Psi_E\rangle \;,\;\qquad E\in \mathbb{R}\;,
\eea
such that $\{ |\Psi_E\rangle\}_{E \in \mathbb{R}}$ forms an
orthogonal basis, labelled continuously by $E$,  of vectors
of an abstract Hilbert space $\mathcal H$ associated with
$H$. Note that we do not
exclude the case of negative spectrum, but will require, as usual, that
the spectrum of $H$ is real. We also assume that
\bea
\langle \Psi_E| \Psi_{E'}\rangle = \delta(E-E')\;,
\qquad  \langle\Psi_E|\Psi_E\rangle=\infty\;.
\eea
Since the states $|\Psi_E\rangle$ are not
normalizable,
none of them can be chosen as a preferred vector
for building normalized coherent states.
Two kinds of {\it regularized} fiducial
vectors $|\eta_{\bar E} \rangle$ and $|\eta_{\bar E} \rangle_A$
will be constructed by a weighted integral of the states $|\Psi_E\rangle$  on an interval
of energy as follows:
\bea
(a)&&
|\eta_{\bar E}  \rangle = C \int_{E_0}^{E_1}   |\Psi_E\rangle
dE\;;
\qquad   \langle\eta_{\bar E}|\eta_{\bar E} \rangle =1
\quad \Rightarrow \quad
C= \frac{1}{\sqrt{\bar E}}   \;;\label{fid1}\\
(b)&&
|\eta_{\bar E} \rangle_A = C_A \int_{-\infty}^{+\infty}   e^{-A(E-\bar E)^2} |\Psi_E\rangle
dE\;,  \quad A \gg 1\;;
\crcr
&&
_A\langle\eta_{\bar E}|\eta_{\bar E} \rangle_A =1
\quad \Rightarrow \quad
C_A= \left(\frac{2A}{\pi}\right)^{\frac14}\;.
\label{fid2}
\eea
Here $\bar E$ is the mean of the weight function which in the uniform case is $(E_1 - E_0)/2$.
Observe that the new vectors $|\eta_{\bar E} \rangle$ and $|\eta_{\bar E} \rangle_A$ are now normalized to 1.
It is noteworthy that the regularization procedure introduced
in (\ref{fid1}) follows that of Isham and Klauder on coherent states for the Euclidean
group $E(n)$ \cite{klish}.
Our remaining goal is to construct two families of vectors
 using $|\eta_{\bar E}  \rangle$ and
$|\eta_{\bar E}\rangle_A$ satisfying a normalizability condition
and possessing a resolution of identity of the Hilbert
space $\mathcal H$. These two requirements
belong to the set of GK physical axioms \cite{gk} for  coherent states. All GK axioms may not be implemented {\it a priori} for the states
that one can define from $|\eta_{\bar E}  \rangle$ or
$|\eta_{\bar E}\rangle_A$; however, that can be remedied later in the analysis.

The definition of coherent states by applying a unitary
operator on the fiducial state will be considered.
In the following, we will focus on the ordinary canonical
phase space quantization, but the same ideas and appropriate
coherent states can be introduced for different types of systems
using the general group theoretical formalism \cite{perel}.

Let us  assume that an ordinary canonical quantization procedure
leads to the quantum Hamiltonian $H$.
On a $2M$ dimensional phase space,
one introduces a set of canonical commutation relations between
 coordinate operators $[Q_k,P_l]=i\hbar \delta_{kl}$,
$k,l=1,2,\dots,M$.
There exists a special class of phase space unitary operators
defined such that $U(\mathbf{q},\mathbf{p})=
\prod_{\ell=1}^M e^{-\frac{i}{\hbar}q_\ell P_\ell}e^{\frac{i}{\hbar}p_\ell Q_\ell}$,
with $q_\ell$ and $p_\ell$ some real parameters, $\mathbf{q}=\{q_\ell\}$ and $\mathbf{p}=\{p_\ell\}$.
These unitary operators will act on the fiducial states
(\ref{fid1}) and (\ref{fid2}), respectively, as
\bea
&&
|\mathbf{q},\mathbf{p};\bar E\rangle = U(\mathbf{q},\mathbf{p})|\eta_{\bar E} \rangle \label{coh1}\\
&&
|\mathbf{q},\mathbf{p};\bar E\rangle_A = U(\mathbf{q},\mathbf{p})|\eta_{\bar E} \rangle_A
 \label{coh2}
\eea
in order to define new families of states.
These two families of states fulfill the following:
\begin{enumerate}
\item[(i)] Normalization condition
\bea
\langle\mathbf{q},\mathbf{p};\bar E |\mathbf{q},\mathbf{p};\bar E\rangle
 =  \langle \eta_{\bar E} |\eta_{\bar E} \rangle =  1\;;  \qquad
\;
_A \langle\mathbf{q},\mathbf{p};\bar E |\mathbf{q},\mathbf{p};\bar E\rangle _ A =\; _A \langle\eta_{\bar E} |\eta_{\bar E} \rangle_A =1 \;;
\eea
\item[(ii)] Continuity in labels:
As $||\mathbf{q} - \mathbf{q}'||_{\mathbb{R}^M} \to 0$
and  $||\mathbf{p} - \mathbf{p}'||_{\mathbb{R}^M} \to 0$
\bea
&&
||\, |\mathbf{q},\mathbf{p};\bar E\rangle -
|\mathbf{q}',\mathbf{p}';\bar E\rangle ||_{\mathcal{H}}^2 = 2
- 2 \Re\; \langle \mathbf{q},\mathbf{p}; \bar E |\mathbf{q}',\mathbf{p}';\bar E\rangle \to 0\;, \crcr
&&
||\, |\mathbf{q},\mathbf{p};\bar E\rangle_A -
|\mathbf{q}',\mathbf{p}';\bar E\rangle_A ||_{\mathcal{H}}^2 = 2
- 2 \Re\; _A \langle \mathbf{q},\mathbf{p}; \bar E |\mathbf{q}',\mathbf{p}';\bar E\rangle_A \to 0 \;,
\eea
where we used the fact that the unitary operators
are weakly continuous in their labels.

\item[(iii)] Resolution of identity: Introducing the eigenbasis
of the position operators $|x_k\rangle$, such that
$Q_k|x_k\rangle= x_k |x_k\rangle$, we consider a
general tensor product of states $|\mathbf{x}\rangle
= \otimes_{k} |x_k\rangle $, such that, we evaluate
\bea
\langle\mathbf{x}|\mathbf{x}'\rangle &=& \int_{\mathbb{R}^M\times \mathbb{R}^M}  \langle\mathbf{x}|\mathbf{q},\mathbf{p};\bar E\rangle\langle\mathbf{q},\mathbf{p};\bar E | \mathbf{x}'\rangle   \;  \frac{d\mathbf{q}d\mathbf{p}}{K}\;,
\crcr
\langle\mathbf{x}|\mathbf{x}'\rangle &=& \int_{\mathbb{R}^M\times \mathbb{R}^M}  \langle\mathbf{x}|\mathbf{q},\mathbf{p};\bar E\rangle_A\; _A\langle\mathbf{q},\mathbf{p};\bar E | \mathbf{x}'\rangle   \; \frac{d\mathbf{q}d\mathbf{p}}{K}\;,
\eea
for some constant $K$.  Focusing on the first integral, we have
\bea
\langle\mathbf{x}|\mathbf{x}'\rangle
&=&  K^{-1} \int_{\mathbb{R}^M\times \mathbb{R}^M}
\langle\mathbf{x}| U(\mathbf{q},\mathbf{p})
|\eta_{\bar E}\rangle  \langle \eta_{\bar E} | U^{\dag}(\mathbf{q},\mathbf{p}) |\mathbf{x}'\rangle   \; d\mathbf{q}d\mathbf{p} \crcr
&=&
(2\pi \hbar)^{M} K^{-1} \delta^{(M)}(\mathbf{x} -\mathbf{x}' )
\int_{\mathbb{R}^M}
\eta_{\bar E}(\mathbf{x} - \mathbf{q})
(\eta_{\bar E}(\mathbf{x} - \mathbf{q}))^* \; d\mathbf{q}\crcr
&=&
(2\pi \hbar)^{M} K^{-1} \delta^{(M)}(\mathbf{x} -\mathbf{x}' )
\int_{\mathbb{R}^M}
\eta_{\bar E}(\mathbf{q})
(\eta_{\bar E}(\mathbf{q}))^* \; d\mathbf{q}
\eea
as follows from a simple change of variable. Setting $K =
(2\pi \hbar)^{M}$, one gets
\bea
\int_{\mathbb{R}^M\times \mathbb{R}^M}  \langle\mathbf{x}|\mathbf{q},\mathbf{p};\bar E\rangle\langle\mathbf{q},\mathbf{p};\bar E | \mathbf{x}'\rangle   \;\frac{d\mathbf{q}d\mathbf{p}}{(2\pi \hbar)^{M}} =  \langle\mathbf{x}|\mathbf{x}'\rangle
\eea
proving that the states $|\mathbf{q},\mathbf{p};\bar E\rangle$
obey a resolution of the identity.
The proof of the resolution of the identity for the second
set of vectors (\ref{coh2}) can be shown using similar ideas.
Thus, both the families of states (\ref{coh1})
and (\ref{coh2}) are proper coherent states.

\end{enumerate}

It is relevant to investigate also how the rest of the GK  axioms,
viz. the temporal stability of states and the action
identity, can be realized in the present setting.

\begin{enumerate}
\item[(iv)]
The temporal stability of the states will not be
fully satisfied initially, but hold only to an approximation.
We seek some conditions under which
the following states should belong to the same family of states
\bea
|\mathbf{q},\mathbf{p}, \tau; \bar E\rangle = e^{-\frac{i}{\hbar} \tau H} |\mathbf{q},\mathbf{p}; \bar E\rangle \;,\qquad
\;
|\mathbf{q},\mathbf{p},\tau; \bar E\rangle _A = e^{-\frac{i}{\hbar} \tau H} |\mathbf{q},\mathbf{p}; \bar E\rangle _A \;.
\label{gentempstab}
\eea
We focus again on the first type of states and rewrite that expression
as
\bea
e^{-\frac{i}{\hbar} \tau H} |\mathbf{q},\mathbf{p}; \bar E\rangle
 =  U (\mathbf{q},\mathbf{p}) e^{-\frac{i}{\hbar} \tau H} | \eta_{\bar E}\rangle
+ [e^{-\frac{i}{\hbar} \tau H} ,U (\mathbf{q},\mathbf{p})] | \eta_{\bar E}\rangle\;.
\eea
The evolution of the fiducial state, i.e.
$
 e^{-\frac{i}{\hbar} \tau H} | \eta_{\bar E}\rangle
 = C \int_{E_0}^{E_1}  e^{-\frac{i}{\hbar} \tau E} | \Psi_{E}\rangle dE $,
can be decomposed using $E=\bar E + \delta$ such that,
for small $\delta$,
\begin{equation}
 e^{-\frac{i}{\hbar} \tau H} | \eta_{\bar E}\rangle
 = C e^{-\frac{i}{\hbar} \tau \bar E}  \int_{E_0-\bar E}^{E_1-\bar E}  (1 - \frac{i}{\hbar} \tau \delta + O(\delta^2)) | \Psi_{\bar E + \delta }\rangle d\delta
= e^{-\frac{i}{\hbar} \tau \bar E} \left[   | \eta_{\bar E}\rangle
-\frac{i}{\hbar} \tau |O(\Delta E)\rangle \right] ,
\end{equation}
where $\Delta E=E_1 -E_0$ and $|O(\Delta E)\rangle$ symbolizes a vector state of norm
of order $O(\Delta E)$. Thus, one gets
\begin{equation}
e^{-\frac{i}{\hbar} \tau H} |\mathbf{q},\mathbf{p}; \bar E\rangle
 = e^{-\frac{i}{\hbar} \tau \bar E} |\mathbf{q},\mathbf{p}; \bar E\rangle
 - \frac{i}{\hbar} \tau e^{-\frac{i}{\hbar} \tau \bar E}U (\mathbf{q},\mathbf{p})\, | O(\Delta E) \rangle +  [e^{-\frac{i}{\hbar} \tau H} ,U (\mathbf{q},\mathbf{p})]\, | \eta_{\bar E}\rangle\;,
\end{equation}
where the  first term is nothing but the initial coherent states
up to a phase, the second term indicates
a state modification depending on the window of integration
$\Delta E$ and the last term involving the commutator
cannot be further computed without knowing the form of the Hamiltonian.

Focusing now on the second type of states with its infinite
integration range and using an analogous procedure,
it can be inferred that the evolution of the coherent states will be
such that the leading order term will be of the form of the
initial coherent state plus other modifications.
 In short, what we reveal here is that
the coherent states may not be temporally stable, in general,
due to the appearance of the second and third contributions.
However, depending on the system under investigation,
this temporal stability may be satisfied at a certain order of
perturbation of the parameters  of the coherent states.
  In the following, we will discuss particular types of system
for which this is indeed the case.

 \item[(v)] The action identity:  Given (\ref{gentempstab}), 
we look for a set of
new canonical variables $(J,\gamma)$ and $(J_A,\gamma_A)$ satisfying
 the following relations
\bea
&& \langle\,\mathbf{q},\mathbf{p},\tau; \bar E | \, H\,  |\mathbf{q},\mathbf{p},\tau; \bar E\,\rangle  = \omega J \;,\qquad\;\dot \gamma = \omega\;,
 \crcr
&&
_A  \langle\,\mathbf{q},\mathbf{p},\tau; \bar E |\,  H  \,|\mathbf{q},\mathbf{p},\tau; \bar E\,\rangle_A  = \omega_A J_A\;,
\qquad\;
\dot \gamma_A = \omega_A\;.
\eea
By inverting $\mathbf{q}(J,\omega)$, $\mathbf{p}(J,\omega)$,
$\mathbf{q}(J_A,\omega_A)$ and $\mathbf{p}(J_A,\omega_A)$,
 and setting $\omega=1=\omega_A$ from initial convention, the following identities hold
\bea
&&
\langle\,\mathbf{q}(J),\mathbf{p}(J),\tau; \bar E |\,  H\,  |\mathbf{q}(J),\mathbf{p}(J),\tau; \bar E\,\rangle  = \langle\, J, \gamma |\,H\,|J,\gamma\,\rangle=  J \;,\quad \gamma = \tau
\crcr
&&
_A  \langle\,\mathbf{q}(J_A),\mathbf{p}(J_A),\tau; \bar E |\,  H\, |\mathbf{q}(J_A),\mathbf{p}(J_A),\tau; \bar E\,\rangle_A  =\crcr
&&
\langle\, J_A, \gamma_A |\, H\, |J_A,\gamma_A\,\rangle =  J_A
\;,\; \gamma_A = \tau.
\eea
Here $|J,\gamma\rangle$ and $|J_A,\gamma_A\rangle$ are viewed as
new labels for the coherent states. This statement depends on the form
of the Hamiltonian and will be discussed for particular examples.

\end{enumerate}

The above basic GK ingredients (i)-(iii) will be
the requirements that the coherent states built in
the sequel should satisfy. We have already shown that
the axiom (iv) will be only valid at a certain parameter
regime whereas (v) will be checked case by case.

\section{Applications}
\label{sect:apps}

Applications of the above formal construction
of coherent states are now provided.
We first discuss in length the case of the free particle and,
in a streamlined analysis,
we study the particle in an inverted
harmonic potential. The latter system has been studied
for years for different purposes \cite{bart}, and recently
seems to have a revival due to the interest of pseudo-bosons \cite{trifo}\cite{gov}\cite{baga}. 
These types of continuous-spectrum systems
 have non-normalizable eigenvectors, and we will illustrate how to construct coherent states for these systems. 

\subsection{Free particle coherent states}
\label{subsec:free}

\noindent{\bf Spectrum and fiducial vector regularization -}
One of the most simple examples for which one can  test
the above ideas is the motion of free particle on a straight
line $\mathbb{R}$.  After canonical quantization and setting $[Q,P]=i\hbar$ for $Q$ and $P$ self-adjoint,
the system can be described by the quantum Hamiltonian given by
\bea
H = \frac{1}{2m} P^2 = \frac{-\hbar^2 }{2m} \frac{d^2}{dx^2}\;.
\label{freeham}
\eea
The Hilbert space of the system is simply given by
the complex span of the position operator $Q$ eigenbasis
denoted by $|x\rangle$.
$H$ admits the following set of eigenvectors and  eigenfuntions
\bea
&&
H |\Psi_k \rangle = \frac{\hbar^2 k^2}{2m}  |\Psi_k \rangle\;, \crcr
&&
\langle x |\Psi_k \rangle  =\Psi_k(x) = \frac{1}{\sqrt{2\pi}} e^{i k x }\;,
\qquad
\langle x |H |\Psi_k \rangle =  \frac{\hbar^2 k^2}{2m} \Psi_k(x)\;.
\eea
The properties of the states $|\Psi_{ k}\rangle$ are direct: they
are not normalizable and they obey
\bea
\langle\Psi_{k} |\Psi_{k'}\rangle = \delta(k'-k)\; \;.
\eea
Thus, for any $k$, $|\Psi_{k}\rangle $ cannot be
chosen as a fiducial vector for defining
coherent states for the system.

The next stage is to define two new functions,
such that for $\bar k\equiv \bar k(k_0,k_1) \in (k_0,k_1)$, $k_0\neq k_1$, and for $A\gg 1$,
\bea
(a)&&
\Psi_{\bar k}(x) = C\int_{k_0}^{k_1} \Psi_k(x) dk
=  \frac{ C}{i \sqrt{2\pi}x}
[e^{i k_1 x } - e^{i k_0 x }]\;,
\crcr
(b)&&
\Psi_{\bar k;A}(x) = C_A\int_{-\infty}^{\infty} e^{-A(k-\bar k)^2}
\Psi_k(x) dk
= \frac{C_A}{\sqrt{2A}}
e^{- \frac{x^2}{4A} + i \bar k x }\;.
\eea
For any $\bar k\in (k_0,k_1)$, we compute the norm of the
first function $(a)$:
\bea
&&
\int_{-\infty}^{\infty}
\Psi^*_{\bar k}(x)\Psi_{\bar k'}(x) dx =
C^2 (k_1 -k_0) = 1\; \qquad
\Leftrightarrow \qquad
C = \frac{1}{\sqrt{ k_1 -k_0}}
\eea
The overlap becomes a finite quantity and is normalized when
$C$ is fixed appropriately as done above.
On the other hand, for the second type of state $(b)$, we have
\bea
\int_{-\infty}^{\infty} (\Psi_{\bar k;A}(x))^* \Psi_{\bar k;A}(x)\, dx =
 C^2_A \sqrt{\frac{\pi}{2A}} = 1 \qquad
\Leftrightarrow \qquad
C_A = \left(\frac{2A}{\pi}\right)^{\frac14}\;.
\label{canorm}
\eea
Thus these states are normalized.

Two  vector states $|\Psi_{\bar k} \rangle$ and
$|\Psi_{\bar k} \rangle_A$ can be defined from
 $ \Psi_{\bar k}(x) = \langle x |\Psi_{\bar k} \rangle $ and
$\Psi_{\bar k;A}(x) = \langle x  |\Psi_{\bar k} \rangle_A $, respectively,
\bea
|\Psi_{\bar k} \rangle= C\int_{k_0}^{k_1}|\Psi_{ k} \rangle \; dk
\;,\qquad \;
|\Psi_{\bar k} \rangle_A = C_A\int_{-\infty}^{\infty}e^{-A(k-\bar k)^2} |\Psi_{ k} \rangle \; dk\;.
\eea
These will be considered as our normalized fiducial vectors.

\medskip

\noindent{\bf Coherent states and GK axioms -}
Introducing the unitary operator
$U(q,p) = e^{-\frac{i}{\hbar}q P} e^{\frac{i}{\hbar} p Q}$, $q\in \mathbb{R}$,
$p\in \mathbb{R}$, we define two families of states such that
\bea
(a)&&
|q,p;\bar k \rangle = U(q,p) |\Psi_{\bar k} \rangle \;,
\quad \langle x|q,p;\bar k \rangle  =\Phi_{\bar k}(q,p;x) = e^{\frac{i}{\hbar}  p (x- q)}\Psi_{\bar k}( x-q)\;,
\label{cohkk}\\
(b)&&
|q,p;\bar k \rangle_A = U(q,p) |\Psi_{\bar k} \rangle_A\;,
\quad  \langle x|q,p;\bar k \rangle_A  =\Phi_{\bar k;A}(q,p;x)
=e^{\frac{i}{\hbar}  p (x- q)}\Psi_{\bar k;A}( x-q).
\label{cohA}
\eea

Let us check the GK axioms of coherent states for
the set of states $\{|q,p;\bar k\rangle\}_{q,p\in \mathbb{R}}$
using the family $\{\Phi_{\bar k}(q,p,x)\}_{q,p\in \mathbb{R}}$.

\begin{enumerate}
\item[(i)] The continuity in labels $q,p$ is obvious.

\item[(ii)] The normalizability condition  can be
checked since it corresponds to the condition that\footnote{
We will use henceforth the compact notation $|\Psi_{\bar k; \bullet}\rangle$ for both kinds of states $\bullet \in \{\emptyset, A\}$ when the statement is valid in both situations.}
$\Phi_{\bar k; \bullet}(q,p,x) \in L^{2}(\mathbb{R},dx)$ is of norm $1$,
the symbol $\bullet \in \{\emptyset, A\}$. We have
\begin{equation}
_\bullet \langle q,p;\bar k|q,p;\bar k\rangle _\bullet =
\int_{-\infty}^{\infty}
(\Phi_{\bar k;\bullet }(q,p,x))^{*} \Phi_{\bar k;\bullet }(q,p,x) \;dx
= \int_{-\infty}^{\infty}
 (\Psi_{\bar k;\bullet }( x-q))^* \Psi_{\bar k;\bullet }( x-q) \;dx
\label{normaliz}
\end{equation}
which, by a change of variable $\tilde x =x-q$, reduces
to the earlier calculations on the $L^2$-normalizability of
$\Psi_{\bar k;\bullet}(x)$.
This result can be naturally seen as
a consequence of the unitarity of the operator $U(q,p)$.

\item[(iii)] The resolution of the identity has to be
verified:
\bea
&&
\int_{\mathbb{R}^2} \;  \langle x |q,p;\bar k\rangle _\bullet \; _\bullet \langle q,p;\bar k|x'\rangle
\frac{dpdq }{2\pi \hbar }=
\int_{\mathbb{R}^2}  \Phi_{\bar k;\bullet}(q,p,x)(\Phi_{\bar k;\bullet}(q,p,x'))^{*}\; \frac{dpdq }{2\pi \hbar }
  \crcr
&&
= \int_{-\infty}^{\infty}\left[\int_{-\infty}^{\infty}
 \Psi_{\bar k;\bullet}( x-q)
\Psi^*_{\bar k;\bullet}( x'-q) dq\right] e^{  \frac{i}{\hbar}p(x - x') } \frac{dp }{2\pi \hbar } = \delta(x-x') =\langle x |x'\rangle \,,
\eea
where we use the fact that the integrations in $q$
and $x$ are similar to compute the term
as performed in (\ref{normaliz}).
Hence $|q,p;\bar k\rangle$ resolves the unity of
  $L^2(\mathbb{R},dx)$.

\item[(iv)] The temporal stability condition will be satisfied
only at the leading order of a small parameter depending on the
two types of coherent states.  However, temporal stability can be
recovered for expectation values as the last step in a calculation.

Let us introduce states endowed with another evolution parameter $\tau$ such that
\bea
|q,p,\tau;\bar k\rangle_{\bullet} = e^{-\frac{i}{\hbar} \tau H} |q,p;\bar k\rangle _{\bullet} \;,
\label{tempstab}
\eea
then of course
\bea
e^{-\frac{i}{\hbar} t H}|q,p,\tau;\bar k\rangle_{\bullet}  = |q,p,\tau+ t;\bar k\rangle_{\bullet} \;.
\eea
Hence the new state $|q,p,\tau;\bar k\rangle_{\bullet} $ implemented with the parameter
$\tau$ is stable under time evolution.
Nevertheless, we could have asked more from our coherent
state and have required that without introducing another parameter,
$e^{-\frac{i}{\hbar} \tau H} |q,p;\bar k\rangle _{\bullet}$
evolves within the same family of states $\{|q,p;\bar k\rangle _{\bullet}\}_{(q,p)\in \mathbb{R}^2}$. This condition should be
considered as the true meaning of temporal stability
under time evolution.The later statement and further consequences
have been also developed in \cite{kl7,rok} (and a way to generate
such states considered in \cite{yaho}). 
 We seek conditions under which
this statement holds.

For any coherent state described above, we have
\begin{equation}
e^{-\frac{i}{\hbar} \tau H} |q,p;\bar k\rangle_{\bullet}
= e^{-\frac{i}{\hbar}q P}
e^{\frac{i}{\hbar} \tau \frac{p^2}{2m}  }
e^{ -\frac{i}{m\hbar} \tau  p P   }
 e^{ \frac{i}{\hbar} p Q  }
|\Psi_{\bar k};\tau\rangle_{\bullet}
=
e^{\frac{i}{\hbar} \tau \frac{p^2}{2m}  }
 e^{-\frac{i}{\hbar}(q + \frac{p}{m} \tau )  P}
 e^{ \frac{i}{\hbar} p Q  }
|\Psi_{\bar k};\tau\rangle _{\bullet} \;,
\end{equation}
where $|\Psi_{\bar k};\tau\rangle _{\bullet} =  e^{-\frac{i}{\hbar} \tau H}   |\Psi_{\bar k}\rangle_{\bullet}  $.

Computing the first type of shifted state $|\Psi_{\bar k};\tau\rangle$,
introducing $\tilde k(k_1,k_0)$, a function to be specified later,
$\delta = k - \tilde k$ and at leading orders in $\delta$, one finds
\bea
&&
\langle x|\Psi_{\bar k};\tau\rangle =
e^{-i  \frac{\hbar \tau}{2m}\tilde k^2 } \langle x|\Psi_{\bar k} \rangle
\label{finpsi}
-\frac{C}{\sqrt{2\pi}} \, (i\tilde k\frac{\hbar \tau}{m})
e^{-i \frac{\hbar \tau}{2m} \tilde k^2 }
e^{i \tilde k  x}
\int_{k_0-\tilde k}^{k_1-\tilde k}
  \delta
e^{i \delta  x}  d\delta \\
&&
+ \frac{C}{\sqrt{2\pi}} e^{-i \frac{\hbar \tau}{2m} \tilde k^2 }e^{i \tilde k  x}\int_{k_0-\tilde k}^{k_1-\tilde k}  O(\delta^2)
e^{i \delta  x}  d\delta \;.
\nonumber
\eea
Defining $\Delta =( k_1 - k_0)/2$ and $\ell=(k_0+k_1)/2 - \tilde k$, and
using the new variable  $\tilde \delta = \delta - \ell$,
 we can re-express the second term integral above as
\bea
&&
e^{i \tilde  k  x}
\int_{k_0-\tilde k}^{k_1-\tilde k}
  \delta
e^{i \delta  x}  d\delta
 =
2i e^{i (\tilde  k +\ell ) x}
\left[
  \frac{ 1}{x^2} \sin( \Delta x)
-
\frac{ 1}{x} \, \Delta \cos( \Delta x)
-  \frac{ i\ell}{ x}
\sin( \Delta x)
 \right] \;.
\nonumber
\eea
Let us consider the vector $|\Pi_\Delta\rangle$ corresponding to
the last quantity and evaluate its norm as
\bea
\int_{-\infty}^{\infty} (\Pi_\Delta(x))^* \Pi_\Delta(x) dx
 = c\,\ell^2 \Delta + O(\Delta^2)\;,
\eea
where $c$ is some constant.
Thus $|\Pi_\Delta\rangle$ is in the Hilbert space
and its  norm is of order $\Delta^{\frac{1}{2}}$.
Introducing this result into the expression (\ref{finpsi}) yields
\bea
&&
\langle x|\Psi_{\bar  k};\tau\rangle =
e^{-i  \frac{\hbar \tau}{2m}\tilde k^2 }\langle x|\Psi_{\bar  k} \rangle
-\frac{i}{2\sqrt{\pi\Delta}} \frac{\hbar \tau \tilde  k}{m} e^{-i  \frac{\hbar \tau}{2m}\tilde k^2 }
\left(
 \langle x| \Pi_\Delta \rangle +  \langle x|  O(\Delta) \rangle \right),
\eea
where the notation $| O(\Delta^n) \rangle$ stands for
a state of norm $O(\Delta^{n})$.
Thus all remainder states are of norm at most  $O(\Delta^{\frac12})$.
By expanding further $e^{i\tau   (2 \bar k\delta  +\delta ^2)} $
in $\delta$, it can also be checked that higher order terms integrated
are of order of magnitude $O(\Delta^{\frac12 + c} )$, $c\geq 0$.
Thus, without any further assumption, the temporal stability condition
is clearly broken at leading order by the term $| \Pi_\Delta \rangle$
with order of magnitude $O(1)$ with respect to the
window of integration $\sqrt{k_1-k_0}$. Nevertheless,
two specific cases of interest may occur:
(1) $0< \Delta\leq \ell \ll 1$, i.e. if the position barycenter is closed to $\tilde k$ (which is the only remaining parameter in the present context and which can be tuned to fulfill that
condition), then we infer that the norm of
$||\,C| \Pi_\Delta \rangle||^2 = O(\ell^2) \ll O(1) = ||\,|\Psi_{\bar k}\rangle||^2$;
(2) $\ell =0$ therefore $||\,C| \Pi_\Delta \rangle||^2 =
O(\Delta) \ll O(1) = ||\,|\Psi_{\bar k}\rangle|$.
Thus, under these assumptions,  we write in shorthand notations
\bea
|\Psi_{\bar k};\tau\rangle  = e^{-i  \frac{\hbar \tau}{2m}\tilde k^2 }|\Psi_{\bar k} \rangle
 + |O(\ell)\rangle \qquad \text{or}\qquad
|\Psi_{\bar k};\tau\rangle  = e^{-i  \frac{\hbar \tau}{2m}\tilde k^2 } |\Psi_{\bar k} \rangle
 + |O(\Delta)\rangle\;,
\eea
such that the evolution of the coherent states is given by
\bea
e^{-\frac{i}{\hbar} \tau H} |q,p;\bar k\rangle   = e^{\frac{i}{\hbar} \tau (\frac{p^2}{2m} -\frac{\hbar^2 \tilde k^2}{2m} )} |q+ \textstyle\frac{p}{m} \tau,p;\bar k\rangle  +(\, |O(\ell)\rangle \leftrightarrow |O(\Delta)\rangle\,)
\label{stab1}
\eea
%{\colred I THINK THE SIGN OF $\tau$ IS WRONG ABOVE.}
with indeed a clear  physical meaning: a given
coherent state evolves within the same family of states
with parameter $q$ translated to $q+  \textstyle\frac{p}{m} \tau$
at order $O(\ell)$ or at order $O(\Delta)$.

For the second type of coherent states,  we calculate
the evolution of the fiducial state in the same manner
as done earlier setting for this time $\tilde k = \bar k$.
We find
\bea
|\Psi_{\bar k};\tau\rangle_A
=  e^{-i  \frac{\hbar \tau}{2m}\bar k^2 } |\Psi_{\bar k}\rangle_A
+(-i \bar k \frac{\tau\hbar}{m} ) e^{-i  \frac{\hbar \tau}{2m}\bar k^2 }
 \frac{1}{A} Q  |\Psi_{\bar k}\rangle_A + \dots
\eea
A calculation establishes that the norm of the remainder state
is
\bea
 \left( \,  _A \langle \Psi_{\bar k}  | (\bar k \frac{\tau\hbar}{m} ) e^{+i  \frac{\hbar \tau}{2m}\bar k^2 }
 \frac{1}{A} Q \right)
\left( (\bar k \frac{\tau\hbar}{m} ) e^{-i  \frac{\hbar \tau}{2m}\bar k^2 }
 \frac{1}{A} Q  |\Psi_{\bar k}\rangle_A \right)
= \frac{1}{A}  \left(\frac{\tau \hbar}{m}\bar k \right)^2
\eea
which is of order $O(1/A)$.  Hence, we have
\bea
e^{-\frac{i}{\hbar} \tau H} |q,p;\bar k\rangle _A  =
e^{\frac{i}{\hbar} \tau (\frac{p^2}{2m} -\frac{\hbar^2 \tilde k^2}{2m}  )}  |q+ \textstyle\frac{p}{m} \tau,p;\bar k\rangle_A  + |O(\textstyle A^{-\frac{1}{2}})\rangle\;.
\label{stab2}
\eea
In conclusion, this family of coherent states is stable under time evolution at the dominant order, for $A$ sufficiently large.

Note that in any situation, interestingly, the inessential phase cancels exactly
for $p^2 = \hbar^2 \tilde k^2 $, namely, when
the classical Hamiltonian picks the value of the quantum
value $\hbar^2 \tilde k^2$.

\item[(v)] The action identity axiom can be examined
by evaluating the mean value of the Hamiltonian.

The coherent states of the first type provide the Hamiltonian mean value
\bea
&&
\langle q,p,\tau;\bar k| H  |q,p,\tau;\bar k\rangle
= \frac{1}{2m}
\langle \Psi_{\bar k} |  (P + p)^2  |\Psi_{\bar k}\rangle \crcr
&&
=    \frac{1}{2m} \left[ \frac13 C^2\hbar^2[k_1^3 -k_0^3]
+  \hbar p C^2 [k_1^2 -k_0^2]  + p^2  \right]= \omega J_{\bar k}(p) \;.
\eea
 The goal is to find
the functions $ p(J, \omega)$ and $q(J, \omega)$ where $J$
will play the role of an action variable and $\omega$ is associated with 
$\gamma$ which will play the role of an angle variable  canonically
 conjugate to $J$.
This can be achieved by considering
\bea
&&
 p_\pm(J,\omega) = - \frac12\hbar C^2 [k_1^2 -k_0^2] \pm
\sqrt{2m\omega J_{\bar k}( p)
-\frac13\hbar^2C^2 [k_1^3 -k_0^3]
 +\frac14 [\hbar C^2(k_1^2 -k_0^2)]^2 }  \;\crcr
&&\arctan \left(\frac {p}{q}\right) = \omega\; \qquad \Rightarrow
\qquad
q_{\pm} (J,\omega)= p_{\pm}(J,\omega) \cot \omega\;.
\label{actangle}
\eea
Hence, setting $\omega=1$  (omitting henceforth the
dependence in $\omega$) and
choosing the relevant root $p_+(J)=p_+(J,\omega=1)$,
the action identity is given by  $\gamma=\tau$ and
\begin{equation}
\langle \,q_{+} (J), p_{+}(J),\tau; \bar k |\,H \,| q_{+} (J),  p_+(J),\tau; \bar k \,\rangle =  J\;.
\label{eqj}
\end{equation}
Meanwhile, the same calculation for the second type of coherent states
yields
\bea
\frac{\hbar^2 \bar k^2}{2m}  +
\frac{\hbar^2}{2m}  \frac{1}{4A}
+\frac{\hbar p   \bar k }{m}
+\frac{p^2}{2m} = \omega_A J_A \;
\eea
with a similar equation for $q$ as in (\ref{actangle}).
Again,  we  invert $p$ in terms of $J_A$ and find
\begin{equation}
p_\pm (J_A,\omega_A) =- \hbar\bar k \pm
\sqrt{ 2m \omega_A J_A  - \hbar \frac{1}{4A} }\;,
\end{equation}
such that the action identity reads, choosing $\omega_A=1$,
 the root $p_+(J_A)= p_+(J_A,\omega_A=1)$,  and $\gamma=\tau$,
\begin{equation}
_{A}\langle\, q_{+} (J_A) ,p_+ (J_A),\tau;\bar k|\, H\,  |q_{+}(J_A),p(J_A),\tau;\bar k\,\rangle _{A}  =  J_A\;.
\label{eqj2}
\end{equation}
 Note that, even though $\tau$ appears in both 
\eqref{eqj} and \eqref{eqj2}, $J$ and $J_A$ do not depend
on $\tau=\gamma$ because, in these equations, that 
dependence explicitly vanishes (complex conjugate phases). Thus $J$ and $J_A$ are
independent and canonically conjugated to $\gamma=\tau$.

\end{enumerate}

\noindent{\bf Saturation of the uncertainty relation -}
The main statistical properties of the set
of states $|q,p,\tau; \bar k\rangle$ cannot be derived
since the state $|\Psi_{\bar k}\rangle$ does not
belong in fact to the domain of $Q$.
Nevertheless, for the second type of coherent states,
one can investigate these properties.
We will not undertake such a statistical analysis in
this paper, however, it is significant to ask
for the saturation of the Heisenberg uncertainty
relation for the type of states studied here.
This is the purpose of this paragraph.

After some algebra, the following identities hold
\bea
&&
\langle Q \rangle = q \;,  \qquad
\langle  Q^2 \rangle = A + q^2 \;,\qquad
(\Delta Q)^2 = A\;, \crcr
&&
\langle P \rangle =  \hbar \bar k + p \;,\;\;
\langle P^2 \rangle = (\hbar \bar k + p)^2
 + \frac{1}{4A}\hbar^2 \;,\;\;
(\Delta P)^2 =  \frac{1}{4A}\hbar^2\; .
\eea
Hence,
\bea
\Delta Q \Delta P  = \frac{\hbar}{2}
\eea
emphasizing the fact that the coherent states defined by (\ref{cohA})
saturate the Heisenberg uncertainty relation.

\noindent{\bf The limit $\Delta \to 0$ and $A \to \infty$ -}
The limits $k_1 \to k_0$ and $A \to \infty$ for the two kinds of coherent states
are of interest. The construction at these limit
situations rests on the following types of states:
\bea
(a)&&
|\Psi_{\bar 0} \rangle = \lim_{k_1 \to k_0} |\Psi_{\bar k} \rangle=
\lim_{k_1 \to k_0} \frac{1}{\sqrt{k_1-k_0}}\int_{k_0}^{k_1}|\Psi_{ k} \rangle \; dk \to 0\;;
\crcr
(b)&&
 |\Psi_{\bar k} \rangle_\infty = \lim_{A \to \infty} |\Psi_{\bar k} \rangle_A =
\lim_{A \to \infty} \left(\frac{2A}{\pi}\right)^{\frac14}
\int_{-\infty}^{\infty} e^{-A(k-\bar k)^2} |\Psi_{ k} \rangle \; dk  \to 0\,.
\eea
Hence, for this limit
the entire coherent state construction is vacuous, as their inner product with $\langle x |$ confirms.
However this does not mean that the limits
$\Delta \to 0$ and $A \to \infty$ are without interest.
In fact, such limits should be performed only {\it after}
taking expectation values. For a general operator $T$, the
following expectations are finite
\bea
\lim_{k_1\to k_0 } \langle q,p; \bar k| T| q,p; \bar k\rangle \;,
\qquad
\lim_{A\to \infty }\;  _A\langle q,p; \bar k| T| q,p; \bar k\rangle_A\;.
\eea
For instance, evaluating the expectations of the identity,
one finds
\bea
&&
\lim_{k_1\to k_0 } \langle q,p; \bar k| q,p; \bar k\rangle
 = \lim_{k_1\to k_0 } \langle \Psi_{\bar k}| \Psi_{\bar k} \rangle =1 \;,\crcr
&&
\lim_{A\to \infty }\;  _A\langle q,p; \bar k|  q,p; \bar k\rangle_A
=\lim_{A\to \infty }\;  _A\langle \Psi_{\bar k}|   \Psi_{\bar k}\rangle_A=1\;,
\crcr&&
\lim_{k_1\to k_0 } \langle q,p; \bar k|H| q,p; \bar k\rangle  =
\frac{p^2}{2m}  \;,\crcr
&&
\lim_{A\to \infty }\;  _A\langle q,p; \bar k| H| q,p; \bar k\rangle_A
=\frac{ (\hbar \bar k + p)^2}{2m}\;.
\eea

\subsection{Inverted harmonic oscillator coherent states}
\label{subsec:harm}

\noindent{\bf Eigenfunctions and states -}
Let us consider the quantum Hamiltonian
\bea
H = \frac{1}{2m} P^2 - \frac{1}{2}m\omega^2 Q^2
 =  \frac{1}{2m}( -\hbar^2 \partial_{x}^2 - m^2\omega^2 x^2)\;,
\label{harm}
\eea
where $[Q,P]=i\hbar$.  This Hamiltonian has been previously
investigated from different perspectives \cite{bart}\cite{jannus}\cite{flo}.
Our goal is to show that the procedure defined above
allows us to define coherent states for this system
with non-normalizable states with an infinite continuous spectrum.
For simplicity in our developments, we will use
notations of \cite{bart}, such that $2m=1=\hbar= \omega$.
The resulting eigenvalue problem
\bea
&&
H \psi_E = E \psi_E \qquad  \Leftrightarrow \qquad
-( \partial^2_x + \frac{1}{4} x^2) \psi_E =
 E \psi_E\;
\label{hamoinv}
\eea
finds solutions  in terms of the parabolic cylinder
functions \cite{abram}
\begin{equation}
\Psi_{E,1}(x) = \,  D_{  -\frac12(1+iE ) }
\left(
e^{\frac14 i\pi }x  \right)
\;,\quad
\Psi_{E,2}(x)  =
\, D_{ -\frac12(1-iE ) }\left(
e^{\frac34 i\pi } x \right) .
\end{equation}
Both of these eigenfunctions are non-normalizable and the system is doubly degenerate.
(See the Appendix for basic properties of $D_\nu(x)$ and other
facts about the Hamiltonian (\ref{harm}).)
Without loss of generality, we will focus on $\Psi_{E,1}(x) $. The latter function
can be decomposed in a linear combination (with non-trivial, energy-dependent coefficients)
of a real and an imaginary part written in terms of $\psi_{E,\pm}(x)=$ $ C_\pm W( E ,\pm x)$ which are
again eigenfunctions of the same operator \cite{abram}.
The constants $C_{\pm}$ can be fixed by the orthogonality
condition (proofs of the following statements
can be found in \cite{bart} or in \cite{abram})
\bea
\int_{-\infty}^{\infty} \psi_{E,s}(x) \psi_{E',s'}(x)\, dx
= K_{s,E} \,\delta_{s,s'}\delta(E-E')\;, \qquad
s,s'=\pm\;, 
\eea
where $K_{s,E}$ is some constant. 
The functions $\psi_{E,s}(x)$ are however non-normalizable:
for $x \gg E$, the following approximations hold \cite{abram}:
\bea
&&
W( E , x) \sim \sqrt{\frac{2\kappa}{x}} \cos\left[\frac14 x^2 + E \log x + \frac{1}{4}\pi + \phi(E) \right] ,\crcr
&&
W(E, -x) \sim \sqrt{\frac{2}{ \kappa x }} \sin\left[\frac14 x^2 + E \log x + \frac{1}{4}\pi + \phi(E) \right], \crcr
&&
\kappa = (1+ e^{-2\pi E})^{\frac12} - e^{-\pi E} \;, \qquad
\phi(E) = \arg \Gamma[\frac12 - iE]\;.
\eea

In the following, we will concentrate on the solution
$\psi_{E}(x)\equiv\psi_{E,+}(x)$, such that the following orthogonality relation will be
relevant
\bea
\int_{-\infty}^{\infty} \psi_{E}(x)\psi_{E'}(x)\,dx = \delta(E-E')\;;
\eea
thus, this will fix
$
C_+ =C_0= \left(2\pi (1+e^{-2\pi E})\right)^{-\frac12} \;.
$

We introduce the state family
$|\psi_{E,+}\rangle$
such that in the $|x\rangle$ representation
we have $\langle x|\psi_{E,+}\rangle$ $= \psi_{E,+}(x)$, $\langle\psi_{E}|\psi_{E'}\rangle$ $= \delta(E-E')$
and $\langle\psi_{E}|\psi_{E}\rangle=\infty$.

\medskip

\noindent{\bf Fiducial vector regularization and
coherent states -}
The regularization of the fiducial vector can be done according
to our prescription as
\bea
&&
| \Psi_{\bar E} \rangle  = C_A \int_{-\infty}^{\infty}
e^{-A(E - \bar E)^2} |\psi_{E}\rangle\,
dE \;,\crcr
&& \langle x | \Psi_{\bar E} \rangle  =C_A  \int_{-\infty}^{\infty}
e^{-A(E - \bar E)^2} \psi_{E}(x) dx =
C_A C_0 \int_{-\infty}^{\infty}
e^{-A(E - \bar E)^2}  W(E,x)
dE\;,
\eea
where $C_A$ is a normalization constant fixed such that the state $| \Psi_{\bar E} \rangle  $ is normalized  in the same way as (\ref{canorm}).

We can check that the set of states defined such that
\bea
&&
|q,p; \bar E \rangle= U(q,p)  | \Psi_{\bar E} \rangle  =
e^{-i q P }e^{i p Q }| \Psi_{\bar E} \rangle \;, \qquad
\langle x |q,p; \bar E \rangle =
e^{i p (x- q)} \Psi_{\bar E}(x-q)
\eea
are coherent states.

\begin{enumerate}
\item[(i)] and (ii): The continuity in labels being obvious, the normalizability condition is achieved by noticing that
\bea
\langle  q,p; \bar E  |q,p; \bar E \rangle  =
\langle \Psi_{\bar E} | \Psi_{\bar E} \rangle  =1\;.
\eea

\item[(iii)]  Let us determine the resolution of the identity
by evaluating the overlap
\bea
&&
\int_{\mathbb{R}^2}
\langle x|  q,p; \bar E \rangle\langle  q,p; \bar E  | x' \rangle \frac{dqdp}{2\pi\hbar} =
\int_{\mathbb{R}^2 }
e^{\frac{i}{\hbar} p (x- q)} \Psi_{\bar E}(x-q)
e^{-\frac{i}{\hbar} p (x'- q)} \Psi^*_{\bar E}(x'-q)
\frac{dqdp}{2\pi\hbar} \crcr
&& = \delta(x-x')\int_{\mathbb{R} }
 \Psi_{\bar E}(x-q)\Psi_{\bar E}(x-q)dq  =\langle x|x'\rangle\;.
\eea
Thus the states $|  q,p; \bar E \rangle$ satisfy
the resolution of the identity.

\item[(iv)]  As we expect, the temporal stability of the states
will only be satisfied at some approximation:
\bea
e^{-\frac{i}{\hbar} t H }  |  q,p; \bar E \rangle
= e^{\frac{i}{\hbar} t \frac{p^2}{2m}  }
 e^{-\frac{i}{\hbar}(q + \frac{p}{m} t )  P}
 e^{ \frac{i}{\hbar} p Q  }
|\Psi_{\bar E};t\rangle
\eea
where we have to expand $|\Psi_{\bar E};t\rangle  = e^{-\frac{i}{\hbar} t H }  |\Psi_{\bar E}\rangle$ as
\bea
&&
\langle x| \Psi_{\bar E};t\rangle
 = C_A  \int_{-\infty}^{\infty}
e^{-A(E - \bar E)^2} e^{-\frac{i}{\hbar} t E } \psi_{E}(x)\; dE
=  C_A  \int_{-\infty}^{\infty}
e^{-A\delta^2} e^{-\frac{i}{\hbar} t (\bar E + \delta) } \psi_{\bar E + \delta}(x)\; d\delta \crcr
&&
= C_A  e^{-\frac{i}{\hbar} t \bar E } \int_{-\infty}^{\infty}
e^{-A\delta^2} (1 - \frac{i}{\hbar} t  \delta + O(\delta^2)) \psi_{\bar E + \delta}(x)\; d\delta \,.
 \eea
We aim at evaluating the norm of
\bea
\langle x | \Upsilon_A \rangle = \Upsilon_A (x) = C_A \frac{i}{\hbar} t  \int_{-\infty}^{\infty}
e^{-A\delta^2} \delta \;\psi_{\bar E + \delta}(x)\; d\delta \;.
\eea
Using the orthogonality of the functions $\psi_{E}(x)$, one has
\bea
\int_{-\infty}^{\infty} \Upsilon^*_A (x)\Upsilon_A (x) dx
= C^2_A \frac{t^2}{\hbar^2}   \int_{-\infty}^{\infty}
e^{-2A\delta^2} \delta^2 \,d\delta =
\sqrt{\frac{2A}{\pi}} \frac{t^2}{\hbar^2}   \frac{\sqrt{\pi}}{4\sqrt{2} A^{\frac32}} =
 \frac{t^2}{4\hbar^2}\frac{1}{A}\;,
\eea
which implies that $\Upsilon_A$ is again of norm $O(1/\sqrt A)$.
Thus, as claimed, temporal stability will be obeyed at
leading order and broken at $O(1/\sqrt A)$.

\item[(v)]  The action identity requires to find $( J_A,\gamma_A)$ such that, given $H = P^2 -\frac14 Q^2$
in appropriate units, we have
\bea
&&
 \; _A
\langle q,p,t;\bar k| \, H\, | q,p,t;\bar k \rangle _A
 =  \; _A\langle \Psi_{\bar E}|(P+p)^2| \Psi_{\bar E} \rangle _A
-\frac14 \; _A \langle \Psi_{\bar E} |(Q+q)^2| \Psi_{\bar E} \rangle  _A  \crcr
&&=
 \; _A\langle \Psi_{\bar E}|H| \Psi_{\bar E} \rangle _A +
2\; _A\langle \Psi_{\bar E}|(pP-\frac14 q Q) | \Psi_{\bar E} \rangle _A
 + p^2 -\frac14 q^2 \crcr
&&
= 2\hbar p K_1(A,\bar E) - \frac12 q K_2(A,\bar E)   + p^2 -\frac14 q^2= \omega_A J_A\;,
\eea
where $K_{1,2}(A,\bar E) $ are functions, yet unknown, obtained after
integrations. We will use the other action-angle equation in order to fix
$(q,p)$:
\bea
\left\{ \begin{array}{cc}
2\hbar p K_1(A,\bar E) - \frac12 q K_2(A,\bar E)   + p^2 -\frac14 q^2 =
\omega_A J_A \\
\arctan \left(\frac{p}{q}\right) = \omega_A
\end{array} \right.
\eea
A rapid inspection shows that this system admits solutions
in terms of $K_{1,2}(A,\bar E)$
which should be analyzed in order to fix the last GK axiom.

\end{enumerate}

\section{Conclusion}
\label{concl}

We have addressed the construction of coherent states
from non-normalizable fiducial states. The main recipe
that we present here is to integrate  some non-normalizable initial  vector
over a range of energy
labels in order to define
a regularized fiducial vector from which any standard procedure
could apply. We have illustrated two kinds of regularization:
one using a sharp cut-off and one using a smooth one.
It is clear that the set of coherent states determined by the second procedure
has  better behaviour with respect to traditional
properties of coherent states than the former set of coherent states.
 In particular, based on the free particle Hamiltonian example,
and focusing on (i) the temporal stability axiom when the relevant parameters are optimized,
and (ii) the saturation of the Heisenberg uncertainty principle,
we have found that the second class satisfies these properties
whereas the first class fails to obey (i) and is simply
undefined for (ii). ({\bf Remark:} This latter deficiency could be resolved by using a smoothed-out smearing
function with compact support of the type belonging to the test-function space ${\cal D}$, e.g.,
a function proportional to
$\exp[-(k_1-k)^{-2}-(k-k_0)^{-2}]$ for $k_0\le k\le k_1$ and zero otherwise; unfortunately
such a weight function is analytically difficult to deal with.)
The temporal stability axiom at leading order of parameters can be restored for the
first class of coherent states with further assumptions.
Similar remarks can be made for the second system we studied, i.e., 
the inverted harmonic oscillator, thereby generating a new
class of coherent states adapted to this system.

If we restrict our requirements, we note that
both formulations lead to coherent states in the original sense
since they are normalized, continuous in their labels, and possess a traditional resolution of the identity.
Furthermore, the formulation we offer here can be useful
in constructing additional sets of normalizable coherent states on Hilbert
spaces with fiducial vectors drawn from non-normalizable dynamical
eigenvectors.

\section*{Acknowledgements}
Comments by Jan Govaerts are gratefully acknowledged.
JRK thanks the Perimeter Institute, Waterloo, Canada, for its hospitality.
Research at Perimeter Institute is supported by the Government of Canada through
Industry Canada and by the Province of Ontario through the Ministry of Research and Innovation.

\appendix
\section*{Appendix: The inverted harmonic oscillator}
\label{app}

\renewcommand{\theequation}{\Alph{section}A.\arabic{equation}}
\setcounter{equation}{0}

We review some basic properties related
with the so-called parabolic cylinder functions
$D_\nu$ and the negative sign or inverted harmonic oscillator.

By definition, the functions $D_\nu(z)$ are solutions
of the Weber differential equation \cite{abram}\cite{whitt}
\bea
\frac{d^2}{d z^2} \, f(z) + (\nu +\frac12 - \frac{z^2}{4})\, f(z)
= 0\;,
\label{weber}
\eea
with $z \in \mathbb{C}$ and $\nu \in \mathbb{R}$.
In fact, there are two independent solutions of this
second order differential equation: $D_\nu(z)$
and $D_{\nu-1}(i z)$. The general solution of (\ref{weber})
is therefore
\bea
f(z) = C_1D_\nu(z) + C_2 D_{\nu-1}(i z)\;,
\label{solweb}
\eea
where $C_1,C_2\in \mathbb{C}$.

For our purpose, we discuss the spectrum of the Hamiltonian
of the inverted harmonic oscillator \cite{bart}:
\bea
H = \frac{1}{2m} (p^2 - k^2 x^2) = \frac{1}{2m}
((-i\hbar)^2 \partial^2_x - k^2 x^2)
\label{hamilt}
\eea
where $k^2=m^2\omega^2$, namely we want to solve
the eigenvalue problem
\bea
H \psi_E = E \psi_E \qquad  \Leftrightarrow \qquad
( \partial^2_x + \frac{k^2}{\hbar^2} x^2) \psi_E =
 -\frac{2m}{\hbar^2} E \psi_E
\eea
with solution according to the above (\ref{solweb})
\bea
\Psi_E(x) &=& C_1 \,  D_{  -\frac12- i  \frac{m }{\hbar k} E  }
\left(
e^{\frac14 i\pi } \left[ \frac{2k}{\hbar} \right]^{\frac12}  x \right)
 + C_2 \, D_{ -\frac12+ i  \frac{m }{\hbar k } E  }\left(
e^{\frac34 i\pi } \left[ \frac{2k}{\hbar} \right]^{\frac12}  x \right) \crcr
&=& C_1 \,  D_{  -\frac12- i  \frac{E }{\hbar\omega}   }
\left(
e^{\frac14 i\pi } \left[ \frac{2 m\omega }{\hbar} \right]^{\frac12}  x \right)
 + C_2 \, D_{ -\frac12+ i  \frac{E }{\hbar\omega }   }\left(
e^{\frac34 i\pi } \left[ \frac{2 m \omega }{\hbar} \right]^{\frac12}  x \right).
\eea
There is another representation of $D_\nu(z)$ given in
terms of confluent hypergeometric series of the first kind or
Whittaker functions
\bea
U(\nu,z) = D_{-\frac12 -\nu }(z) \;.
\eea
We have the asymptotes for complex $z$,
$z$ large in the sense of Watson, i.e.  $|z|\gg 1$ and $|\arg z|< \frac12 \pi$
\cite{abram},
\bea
U(\nu,z) \sim e^{-\frac14 z^2 } z^{-\nu - \frac12 }
\left(1- \frac{(\nu+\frac12)(\nu+\frac32)}{2z^2}\right).
\eea
Therefore
\bea
U\left( i\frac{E }{\hbar\omega },
e^{\frac14 i\pi } \left[ \frac{2 m\omega }{\hbar} \right]^{\frac12}  x \right)
 \sim K  e^{-\frac14 i \left[ \frac{2 m\omega }{\hbar} \right]  x^2 }
(e^{\frac14 i\pi } \left[ \frac{2 m\omega }{\hbar} \right]^{\frac12}  x )^{-i\frac{E }{\hbar\omega }  - \frac12 }\;.
\label{asympt}
\eea
Up to an overall constant, we loosely write, using the principal
branch of logarithm to define complex powers
which should be compatible with $|\arg e^{-\frac14 i\pi}|=\frac\pi4 < \frac12 \pi$,
the absolute square of the amplitude becomes
\bea
&&
\left|U\left( i\frac{E }{\hbar\omega },
e^{\frac14 i\pi } \left[ \frac{2 m\omega }{\hbar} \right]^{\frac12}  x \right)
\right|^2
 \sim K   |x ^{-i\frac{E }{\hbar\omega }  - \frac12 }|^2
 = K |x|^{-1} e^{-i \frac{E }{\hbar\omega } (\ln |x| + i\arg x)}
e^{+i \frac{E }{\hbar\omega } (\ln |x| - i\arg x)} \crcr
&&
\sim K |x|^{-1}
\eea
which is not integrable on the real line $x \in \mathbb{R}$.
This result can be equivalently  derived by the WKB method
from the previous Hamiltonian problem.

Putting  the Hamiltonian (\ref{hamilt}) in the form
\begin{equation}
H = \frac{-k^2}{2m} (x^2 - \frac{1}{k^2}p^2 )\;,
\label{hamo}
\end{equation}
we define
\begin{equation}
\tilde a = \sqrt{\frac{k}{2i\hbar}} (x + \frac{1}{k} p)\;, \quad
 a = \sqrt{\frac{k}{2i\hbar}} (x - \frac{1}{k} p) \;, \qquad
[a,\tilde a] = \frac{k}{2i\hbar} ( i \frac{\hbar}{k} + i\frac{\hbar}{k}  ) = 1 \;.
\end{equation}
Note that using the ordinary scalar product of $L^2(\mathbb{R},dx)$,
$\tilde a$ is clearly not the adjoint of $a$. Indeed, using $x^\dag = x$ and
$p^\dag = p$, one learns that $a^\dag = -ia$ and so
$[a,a^\dag]=0$.  These properties lead to the so-called
algebra of pseudo-bosons  for which two operators
obey
$
[a,b] =1
$
without being adjoint of one another \cite{trifo}\cite{gov}\cite{baga}. Here the discussion
is even  closer to the damped harmonic oscillator
in the sense of \cite{baga}, i.e.,
\bea
&&
H = -\frac{i\hbar\omega}{2}(   \tilde a  a +  a  \tilde a  )
 =- i \hbar \omega (\tilde a  a +\frac12 )\;,
\crcr
&&
H^\dag = i\hbar \omega (a^\dag(\tilde a)^\dag   +\frac12 )
 = - i\hbar \omega ( a \tilde a   -\frac12 )
 = H \;,
\eea
as should be expected from (\ref{hamo}).


\begin{thebibliography}{99}

\bibitem{kl} Klauder J R 1963 {\it J. Math. Phys.} {\bf 4} 1058

\bibitem{kbs} Klauder J R and Skagerstam B-S 1985
``Coherent States'' (Singapore, World Scientific)

\bibitem{perel}
  Perelomov A M 1986
  ``Generalized coherent states and their applications,''
  (Springer, Berlin)

\bibitem{kl2} Klauder J R 1993 {\it Mod. Phys. Lett.} A {\bf 8} 1735

\bibitem{kl3} Klauder J R 1995 {\it Ann. Phys.} NY {\bf 237} 147

\bibitem{kl4} Klauder J R 1996 {\it J. Phys. A: Math. Gen.} {\bf 29} L293

\bibitem{od} Odzijewicz A 1998  {\it Commun. Math. Phys.} {\bf 192} 183

\bibitem{gk} Gazeau J-P  and Klauder J R 1999
{\it J. Phys. A: Math. Gen.} {\bf 32} 123


\bibitem{al}
Ali S T,  Antoine J-P and  Gazeau J-P  2000
``Coherent States, Wavelets, and their Generalizations''
(Springer-Verlag, Berlin)

\bibitem{kl7} Klauder J R ``The current state of coherent states''
Contribution to the 7th ICSSUR Conference June 2001
 arXiv:quant-ph/0110108

\bibitem{klish}
Isham C and Klauder J R 1991 {\it J. Math. Phys.} {\bf 32}  607

\bibitem{kbg}
Ben Geloun J and Klauder J R 2009
{\it J. Phys. A: Math. Theor.} {\bf 42} 375209

\bibitem{new}
 Honarasa G R,   Tavassoly M K, Hatami M and Roknizadeh R 2011
{\it J. Phys. A: Math. Theor.} {\bf  44}  085303


\bibitem{bart}
Barton G 1986 {\it Ann. Phys. } {\bf 166} 322


\bibitem{jannus}
Jannussis A and Skuras E 1986 {\it Lett. Nuovo Cimento}
{\bf 94} 29

\bibitem{flo}
Lo C F  1990 {\it Phys. Rev. A} {\bf 42} 6752


\bibitem{jose}
Capmany J and Fernandez-Pousa  C R
2011 {\it J. Phys. B: At. Mol. Opt. Phys.} {\bf 44}  035506


\bibitem{trifo}
 Trifonov D A 2009 in
``Differential Geometry, Complex Analysis and Mathematical Physics,''
eds. K. Sekigawa et al (W. Scientific, Singapore),
page 241 arXiv: 0902.3744[quant-ph]


%\cite{Govaerts:2009dh}
\bibitem{gov} 
  Govaerts J, Bwayi C. M. and Mattelaer O, 2009
 {\it  J. Phys. A: Math. Theor.} {\bf 42}, 445304 


\bibitem{baga}
Bagarello F 2011
``Dissipation evidence for the quantum damped harmonic oscillator via pseudo-bosons'' arXiv:1106.4638[math-ph]


\bibitem{rok} 
  Roknizadeh R and  Tavassoly M K 2005
 {\it  J. Math. Phys.} {\bf 46} 445304 

\bibitem{yaho}
 Yadollahi F and Tavassoly M K 2011
 {\it  Optics Communications} {\bf 284}  608 



\bibitem{abram}
``Handbook of Mathematical Functions'' 1972
10th edition {\it Appl. Math. Ser.} {\bf 55}
Section 19  Abramowitz A  and  Stegun  I A editors (Dover, NY)

\bibitem{whitt}
Whittaker E T and  Watson G N 1990
``A Course of Modern Analysis'' 4th Ed
(Cambridge University Press, Cambridge)


\end{thebibliography}
\end{document}